# Dispelling the antihydrogen myth

G. Van Hooydonk, Ghent University, Faculty of Sciences, B-9000 Ghent, Belgium

Abstract. *While achiral Bohr atom theory cannot generate Hbar signatures, achiral Heitler-London bond theory can but its Hbar signatures must be detected. We show that the largest spectral signature to probe Hbar is the singlet-triplet splitting of 9,5 eV at $r_0=0,74$ Å, observed in the dihydrogen band spectrum. This large Hbar-signature, overlooked for nearly a century, is confirmed with the observed $H_2$ potential energy curve. Hbar claims by CERN-based collaborations, seemingly important for the fate and future of Hbar, are premature and must be examined critically.*
Pacs: 36.10.-k

## I. Introduction

The CERN Large Hadron Collider (LHC), to be operational in 2008, raises great expectations for particle physics at large and for antimatter- and H̲-physics [1-3] in particular. The long-held view that matter and antimatter are not equally represented in the Universe is based mainly on the failure to detect spectral signatures for antimatter. To find out about its spectral characteristics, attempts to synthesize H̲ started at CERN a few years ago. However, 6 years and hundreds of H̲-papers after the first claims for H̲ mass-production by ATHENA and ATRAP H̲-collaborations [1,2], there is still no evidence that H̲ was trapped and only indirect evidence that H̲ may have been produced [3]. This uncertainty surrounding [1-3] leads to a few remarks.
(i) Until today, ATHENA and ATRAP failed on spectral evidence for H̲, e.g. its 1S-2S term, without which it is impossible to probe the presence of H̲. Since determining the H̲-spectrum is exactly their goal, spectral signatures with which to probe H̲, cannot be given. Claims [1,2] are, to say the least, premature because of the lack of hard evidence for H̲.
(ii) Prior to [1,2], H̲-signatures were found in the line spectrum of atom H [4] and in the band spectrum of molecule $H_2$ [5,6], two spectra available for almost a century. If H̲-signatures show in 2 and 4 particle systems H and $H_2$ [4,5], something very elementary must be wrong with [1,2] and with the so-called matter-antimatter asymmetry in the Universe.
(iii) The Mexican hat curve for natural and stable atom H [4] exposes its chiral fine structure with its 2 wells, separated on the field axis $r=n^2 r_B$ [4]. This Hund-type quartic gives away an intra-atomic phase transition between left- and right-handed hydrogen H ⇆ H̲, with an achiral state $H_a$ at $n=\pi$ in between [4]. If H goes over in H̲ at critical separation $r_c$ in region $r_B < r_c < \infty$, the observed dissociation and/or combination process

$$e^- + p^+ \leftrightarrows H \qquad\qquad (1a)$$

must be refined, knowing electron-proton and positron-antiproton attraction both follow $-e^2/r$. With Mexican hat curve [4], H ($e^-$;$p^+$) is confined to long range $\pi \leq n \leq \infty$; $\underline{H}$ ($e^+$;$p^-$) to short range $0<n<\pi$ or $0<r<r_c$. If so, achiral process (1a) must be adapted to give chiral interaction process

$$e^+ + p^- \leftrightarrows \underline{H} \rightarrow H_a \leftarrow H \leftrightarrows e^- + p^+ \qquad (1b)$$

$$0\leftarrow r \qquad r_B \qquad r_c \qquad r\rightarrow\infty$$

$\underline{H}$-production at CERN [1,2] follows a long-range combination reaction, derived from (1a), i.e.

$$e^+ + p^- \rightarrow \underline{H} \qquad (1c)$$

Although their combination reaction (1c) seems plausible when looking at (1a), it is not consistent with chiral process (1b), which forbids $\underline{H}$ at long range [7]. This may explain difficulties in [1,2] with $e^+$ and $p^-$ beams to make unlikely interaction (1c) happen nevertheless. While $\underline{H}$-signatures (1b) apply for atom hydrogen, we now expand on theoretical $\underline{H}$-signatures in a chemical environment [5], despite the fact that $\underline{H}$-signatures [4-6] are hitherto ignored. Whereas achiral Bohr theory cannot distinguish between different symmetries for H and $\underline{H}$, achiral Heitler-London bond theory generates molecular states with different symmetries for HH, $\underline{HH}$, H$\underline{H}$ and $\underline{H}$H [5]. We prove that the observed splitting for natural dihydrogen must be interpreted as the largest $\underline{H}$-signature ever observed. Since it shows in a band spectrum, known for a century [5], this molecular $\underline{H}$-signature dispels the myths, surrounding antimatter and $\underline{H}$.

The outline is as follows. Theoretical $\underline{H}$-signatures for hydrogen dimers are in Section II. In Section III, the $\underline{H}$-controversy is considered as a controversy on the better of 2 theories to explain splitting in dihydrogen. Results with two theories are in Section IV, while Section V gives supporting evidence for the simpler theory. The conclusion in Section VI favors natural $\underline{H}$-states.

## II. H in atomic Bohr theory and $\underline{H}$ in molecular Heitler-London theory

Bohr's fairly accurate achiral H theory cannot distinguish between H ($e^-$, $p^+$), say in state $\rightarrow$[1], and charge-inverted H ($e^+$, $p^-$), inverted state $\leftarrow$. Charges $e^-$; $p^+$ in H are assigned according observed *long range process* (1a) but this is not conclusive for *short range behavior* (1b). Bohr's energy formula

$$E_n(\rightarrow) \equiv E_n(\leftarrow) = -R_H/n^2 \qquad (2a)$$

where $R_H$ is the Rydberg, is identical for both states. In fact, the same 2-term Hamiltonian

$$\mathbf{H} = \tfrac{1}{2}\mu v^2 - e^2/r \qquad (2b)$$

with $\mu = mM/(m+M)$, applies for both H and $\underline{H}$, all particles having the same positive masses m and M. In atomic Bohr theory, splitting $S_H$ between H and $\underline{H}$ states

$$S_H = \Delta E = E_n(\rightarrow) - E_n(\leftarrow) = 0 \qquad (2c)$$

---
[1] Arrows $\rightarrow$ and $\leftarrow$ indicate that the effect of intra-atomic charge inversion shows on field axis r for both electron-proton and positron-antiproton Coulomb attractions $-e^2/r$. Arrows up $\uparrow$ and down $\downarrow$ refer to spin effects.

is zero. Chiral refinements are needed to expose the differences in (1b) [4]. A test of (2a) with running Rydbergs $R_n=n^2.E_n$ reveals, in a phenomenological way, that $R_H$ is not constant at all, which leads to *atomic* H-signatures [4]. To do justice to Bohr, H was not really an issue in 1913. Only around 1930, antimatter entered the scene with the discovery of positron $e^+$ and with Dirac theory. This explains why also Heitler and London's (HL) 1927 QM solution for bond $H_2$ did not refer to H-states either[2] [5]. Just like Bohr H theory is an *achiral atom theory*, HL $H_2$ theory is an *achiral bond theory*. Although *the line spectrum of atom hydrogen* is constrained by Bohr's $S_H=0$ (2c), it was known for long that *the band spectrum of the dihydrogen bond* shows a large splitting $S_{HH}$ between *2 molecular states with different symmetries*, the lower being a *bound singlet* state, the ground state; the upper an *unbound triplet* state. The 1927 HL explanation for $S_{HH}$ was accepted without questions. At the equilibrium inter-nucleon separation $r_0$, the observed splitting $S_{HH}$, hardly visible at long range [5], is quite large and about twice the $H_2$ dissociation energy $D_e$ of 38500 cm$^{-1}$ or

$$S_{HH}(r_0)=2D_e=77000 \text{ cm}^{-1} = 9,5 \text{ eV} \qquad (2d)$$

HL used Hamiltonian **H** for 2 H atoms a,A; b,B, where a,b are negative leptons and A,B positive nucleons. The 2 H atoms being symmetric as to charges, the 10 terms in Hamiltonian $\mathbf{H}_S$ are

$$\mathbf{H}_S=(\tfrac{1}{2}m_av^2+\tfrac{1}{2}m_bv^2+\tfrac{1}{2}m_Av^2+\tfrac{1}{2}m_Bv^2-e^2/r_{aA}-e^2/r_{bB})+(-e^2/r_{aB}-e^2/r_{bA}+e^2/r_{ab}+e^2/r_{AB})$$
$$=\mathbf{H}_0 + \Delta\mathbf{H} \qquad (2e)$$

(intra-atomic terms give sum $\mathbf{H}_0$; inter-atomic terms $\Delta\mathbf{H}$). Lower + sign in (2e) is conventional: it only reminds that like charge distributions give *inter-nucleon Coulomb repulsion* $+e^2/r_{AB}$.

Since even refined atom theories[3] prescribed degenerate H and H spectra, drastic measures were taken. With charge distribution H (e$^-$; p$^+$), supported by (1a), not only neutral antimatter atom H (e$^+$; p$^-$) was banned from the natural (matter) world but also all H-containing systems. However, physicists still wondered if, amongst others, antimatter species H would obey CPT or not. To find out, mass-produced H [1,2] should give the H spectrum, e.g. its interval 1S-2S. To allow an accurate comparison with H, H interval 1S-2S was already measured within 1,8 parts in $10^{14}$ [8]. However, while all atom theories are ineffective on anti-atom H, HL theory, based on (2e), can deal formally with charge-inversion effects in HH, HH and HH [5]. Its theoretical results for H-states can be tested with experiment, which is the goal we set in this paper [5].

(i) Antimatter bond HH also obeys $\mathbf{H}_S$ (2e) exactly: none of the Coulomb terms in (2e) changes sign [5]. If $\mathbf{H}_S$ (2e) applies for HH, spectra of HH and HH are degenerate, exactly as for H and H (2c). As a result, H̲H̲ was banned from nature, following the general veto on all antimatter. With fields[1], HH is sum state →→; H̲H̲ inverse sum state ←←, giving *2 sum states* (+).

---

[2] A linear combination of $\varphi_I$ and $\varphi_{II}$, wherein coordinates for leptons and nucleons are exchanged, gives symmetric $\varphi_S=\varphi_I+\varphi_{II}$ and antisymmetric $\varphi_A=\varphi_I-\varphi_{II}$. Pauli's principle imposes antisymmetry for lepton spins, which leads to a *singlet* for the ground state of $H_2$ and a *triplet* for the repulsive state (see below).

[3] Also with Dirac bound state H theory, a copy of Sommerfeld's older relativistic H theory, spectra of H and H had to be identical. We do not elaborate here on bound state QED for *atoms* to not distract from H in *molecules*.

(ii) For asymmetrical HH̲ and H̲H, (2e) must be adapted for anti-symmetrical anti-parallel →← and inverted anti-symmetrical anti-parallel ←→, giving *2 difference states* (-). Hamiltonian $\mathbf{H}_A$ is [5]

$\mathbf{H}_A = (½m_av^2 + ½m_bv^2 + ½m_Av^2 + ½m_Bv^2 - e^2/r_{aA} - e^2/r_{bB}) - (-e^2/r_{aB} - e^2/r_{bA} + e^2/r_{ab} + e^2/r_{AB})$

$\qquad = \mathbf{H}_0 - \Delta\mathbf{H}$ \hfill (2f)

wherein lower – sign refers to *inter-nucleon Coulomb attraction* $-e^2/r_{AB}$, in contrast to (2e). While $\mathbf{H}_0$ in both (2e) and (2f) does not vary with separation r between the 2 atomic species, $\Delta\mathbf{H}$ does. Since the 2 Hamiltonians $\mathbf{H}_S$ (2e) and $\mathbf{H}_A$ (2f) describe hydrogen systems with *different discrete symmetries*, the spectra of their sum and difference states are split in function of r exactly by

$\qquad S_\pm(r) = |2\Delta\mathbf{H}|$ \hfill (2g)

*Formal result (2g) is valid without any calculation or any wave function needed* [5]: *it not would not only explain splitting like (2d) in a generic way*[4]; *if (2g), due to intra-atomic charge inversion, were really be at the basis of observed (2d), it would probe the presence of H̲ with a band spectrum, wherein splitting (2d) is observed.*

**III. H̲-controversy: how to explain splitting in dihydrogen**

*III.1 Computational difficulties to account for splitting in 4-particle system dihydrogen*

Following the notation for $S_\pm$, $\mathbf{H}_S = \mathbf{H}_+$ and $\mathbf{H}_A = \mathbf{H}_-$ leads to compact algebraic pair [5]

$\qquad \mathbf{H}_\pm = \mathbf{H}_0 \pm \Delta\mathbf{H}$ \hfill (3a)

Whatever the sign of $S_\pm$, algebraic pair (3a) suffices to explain, *at least conceptually*, why the band spectrum for dihydrogen shows 2 states with *different symmetries*. Unfortunately, it is impossible to conclude from (3a) which is the more stable state, since the functional dependence on r is not known: (3a) and (2g) are simply insoluble, however appealing by their formal simplicity. The same difficulties apply for all 4-particle systems, known to be insoluble almost by definition. Although QM uses only Hamiltonian $\mathbf{H}_+$ without intra-atomic charge inversion, it faces similar difficulties to calculate $S_{HH}$ (2d). In practice, QM proves extremely difficult, hard to generalize and certainly far from transparent, even for the simplest bond of all, dihydrogen [9]. The wave equation with (2e) is only reasonably *soluble* with the BOA (Born-Oppenheimer approximation) [10]. The best approximate QM BOA solution for $H_2$, the simplest bond of all, is due to Wolniewicz [9]. However, to get accurate results for $H_2$ quanta [11], he needed many parameters for optimization and his *best* wave function contains not less than 278 terms [9].

---

[4] Of course, it is possible to generate anti-symmetry in 4-particle systems by changing the positional coordinates of leptons and nucleons (*wave functions*), which is the basis of HL theory with $\mathbf{H}_S$. Using the word *generic* here is justified, since the splitting between $\mathbf{H}_S$ and $\mathbf{H}_A$ states is completely independent of particle coordinates in wave functions.

To get out of the most urgent problem with algebraic Hamiltonian pair (3a), i.e. state stability, its ± signs must be connected *unambiguously* with terms, decisive for state stability [5]. Since nucleons have the greater inertia, classical physics suggests that the term with inter-nucleon separation $r_{AB}$, e.g. $\pm e^2/r_{AB}$, is the more likely to be responsible, if not decisive, for state stability.

*III.2 Born's three bond approximations* [10,12-13] *and the algebraic Hamiltonian pair*

In the BOA [10], nucleons at rest secure that 2 states are described primarily with *inter-nucleon Coulomb interactions* $\pm e^2/r_{AB}$, say $\pm e^2/r$ after all lepton-lepton and lepton-nucleon terms are separated from term $\pm e^2/r$. As a result, the BOA transforms (3a) in a similar algebraic pair

$$\mathbf{H}_{\pm(BOA)} = \mathbf{H}_{0(BOA)} \pm e^2/r \qquad (3b)$$

Its charge symmetric BOA- and anti-symmetric BOA-states (or antiBOA-states)

$$\mathbf{H}_{(BOA)} = \mathbf{H'}_0 + e^2/r \qquad \text{for HH, } \underline{HH} \qquad (3c)$$

$$\mathbf{H}_{(antiBOA)} = \mathbf{H''}_0 - e^2/r \qquad \text{for H}\underline{H}, \underline{H}H \qquad (3d)$$

are connected *unambiguously* with *sum states* HH and $\underline{HH}$ and *difference states* H$\underline{H}$ and $\underline{H}$H. Although a BOA scheme (seemingly) overlooks 9 of 10 terms in the total Hamiltonians (2e) and (2f) [5], it is nevertheless reliable for its 2 states (3b). Since BOA splitting follows Coulomb's law

$$S_{BOA}(r) = |2e^2/r| \qquad (3e)$$

the *repulsive or attractive* character of states (3c)-(3d) is now unambiguously defined. Unlike (3a), (3b) readily quantifies *splitting* $S_{BOA}$ (3e) but its value is constrained by the validity of the BOA. *Of all possible theoretical approximations thus far for splitting in (insoluble) 4-particle system dihydrogen $S_{HH}$ (2d), $S_{BOA}$ (3e) is the only one to provide with an explicit quantitative and extremely simple result.*

Before proceeding, we must find out more about the meaning of the BOA. We therefore discuss all 3 bonding approximations, proposed by Born [10,12-13].

(i) A first remark is that antiBOA (3d), valid exclusively for H$\underline{H}$ and $\underline{H}$H, is not only of classical 19[th] century *ionic* type [5], it is also similar to Born's 2 classical bond approximations [12-13].

(ii) The first and oldest bond approximation by Born and Landé [12], of antiBOA-type (3d), is

$$V(r) = B/r^n - e^2/r \qquad (3f)$$

and appeared many *years before* BOA [10]. The 3[d] and latest by Born and Mayer [13] is similar to (3f) and still of antiBOA-type (3d), although it appeared *years after* BOA [10].

(iii) It is not always realized that BOA (3c) and Born's other bond approximations [12,13] with antiBOA (3d), are *mutually exclusive*. Born nevertheless proposed the three, although he must have known that his *classical* 1[st] and 3[d] [12,13] contradict his 2[d] *non-classical* BOA [10], used in QM.

(iv) Born's oldest classical potential (3f) generalizes the *ionic* Sommerfeld-Kratzer potential [14,15]

$$V(r) = B/r^2 - e^2/r \qquad (3g)$$

since n=2 in (3f) returns (3g). Classical *ionic* potential (3g) is not only useful for *covalent* $H_2$ [15], it also rationalizes the spectral behavior of all bonds between univalent atoms [5,6] and accounts for all observed $H_2$ levels with *spectroscopic accuracy* (errors of 0,015 cm$^{-1}$ [16], smaller than in [9]).

(v) The closed form oscillator form behind (3g) is an *ionic* Kratzer Coulomb oscillator potential

$$V'(r) = \tfrac{1}{2}(e^2/r_0)(1-r_0/r)^2 \qquad (3h)$$

(see also Section V). Kratzer potential (3g)-(3h) is important for many reasons [5,6,15,16]

Having said this on Born-approximations [10, 12-13], we return to (3b), which provides with the stability criterion needed: if one state is relatively *repulsive*, the other is *attractive*, seemingly a trivial result. With classical physics, the energy of the more stable state must lower with decreasing $r_{AB}$, *the inter-nucleon separation*. Since BOA (3c) used in QM, is repulsive in terms of classical physics, it is not the best of choices for the ground state. Therefore, only states obeying antiBOA (3d) are *attractive* where it really matters[5], i.e. in the region $r_0 \leq r \leq \infty$.

As a result, the dihydrogen singlet ground state follows *attractive* antiBOA (3d), exclusively valid for H<u>H</u> and <u>H</u>H, whereas the triplet state follows *repulsive* BOA (3c), exclusively valid for HH and <u>HH</u>. Since the 2 *mutually exclusive states of different symmetry* do not intermix as revealed by splitting $S_{BOA}$ (3e) and by observed $S_{HH}$ (2d), these results are not trivial[6]: they are conclusive for the fate of <u>H</u> and even stand *without any calculation or any wave function*.

Anti-symmetric pair H<u>H</u>; <u>H</u>H further secures that the bond has no permanent dipole moment [5]. Although these qualitative results on the basis of *dynamic symmetries* prove conclusive on the fate of <u>H</u> in natural systems [5] and contradict [1,2], quantitative results are needed in support.

## IV. Largest <u>H</u>-signature ever in nature

Solving the <u>H</u>-problem being equivalent with solving $S_{HH}$ (2d) in dihydrogen, we test concurrent approaches (i) *complex* BOA QM which bans <u>H</u>, and (ii) *conceptually simple* theories (3a)-(3b), which allow <u>H</u>. While in (i) the analytical form of splitting is very complicated [9], splitting with (ii) is extremely simple with $S_{BOA}(r)$ in (3e), with only one Coulomb term $2e^2/r$. This is soluble without any effort, if first principles effects of reduced mass and virial accounted for (see below).

---

[5] This remark is valid unless repulsion can become attraction, which is impossible by definition. Yet, this is exactly the QM procedure, achieved with the intermediary of wave functions [5] (see also footnote 4).

[6] Repulsion, needed to generate the periodic vibrations in combination with attraction, is also anti-symmetric: in the ground state, anti-symmetric state →← gives *attraction*; inverted anti-symmetric state ←→ gives *repulsion* [16]. In either case, anti-symmetry is respected and splitting is avoided accordingly.

*IV.1 Results with QM theory without intra-atomic charge-inversion, i.e. without H̲*

Symmetry based splitting in QM relies on positional coordinates for leptons and nucleons in wave functions $\psi_\pm=\psi_{1,2}\pm\psi_{2,1}$ with lepton spin-symmetry and –antisymmetry[2,4,5]. However, small lepton spin energy effects of order 1 cm$^{-1}$, showing in the H fine structure [4], can never account for a splitting as large as 77000 cm$^{-1}$ (2d). This brings in not transparent, complex QM (with 278 terms for the wave function of simple $H_2$ [9] and BOA [10]), which accounts for the complete molecular band spectrum within the experimental errors of Dabrowski [11] and therefore also for observed $S_{HH}$. *If QM were really reliable for $S_{HH}$, H̲ is superfluous to explain the observed splitting in the $H_2$ band spectrum. However, this solution does not really settle the H̲-problem; it avoids the problem by banning H̲ from the natural world. Although this ad hoc solution for H̲ is accepted in mainstream physics and eventually led to H̲-experiments like [1,2], it denies the subsequent problem that QM cannot deal conclusively with H̲-containing systems* [5,17].

If QM were accurate for HH, it must be as accurate for H̲H̲ and even for HH̲ and H̲H. If QM makes sense, using Wolniewicz's parameters and wave function [9] in a wave equation with **H**$_-$ instead of **H**$_+$, H̲-systems should be described as accurately as HH [9]. Numerous studies on HH̲-interactions and -stability as well as on the HH̲ PEC [5,17] reveal that QM is not unanimous at all on H̲. Since QM is not conclusive at all for H̲-systems, it is less reliable than it seems, which justifies searches for alternative more conclusive theories.

*IV.2 Results with simple bond theories (3a)-(3b) with charge inversion, i.e. with H̲*

H̲-based theories are not only conceptually simpler [5]; they also rationalize the behavior of all systems of interest, e.g. HH, H̲H̲, HH̲ and H̲H, which QM cannot do. Solving (2g) in a simple way is possible with a two-fold Hamiltonian symmetry [5]. Without giving details, splitting (2g) reduces from 4 Coulomb terms to only 2, i.e.

$$S_\pm(r_0) \sim |2e^2/r_{ab}+2e^2/r_{AB}| \approx |4e^2/r_0| \qquad (4a)$$

if, in first approximation, $r_{ab}=r_{AB}$ at $r=r_0$ [5]. Using compact BOA (3e) with only one term, BOA splitting is of very simple *electrostatic, Coulomb ionic nature*. Pending BOA difference $|\mathbf{H'}_0 - \mathbf{H''}_0|$, whereby nucleon kinetic energy is suppressed, and without any standard corrections applying at for $r=r_0$, *Coulomb attraction* in $H_2$ at $r_0=0{,}74$ Å [18] would give

$$S_{BOA}(r_0)=2e^2/r_0=2.116000/0{,}74 =314000 \text{ cm}^{-1} \approx 4S_{HH} \qquad (4b)$$

obviously too large by a factor of 4, compared with observed $S_{HH}$ (2d). By the same argument, the correction factor for (4a) is 8, twice as large. Both results show that splitting with H̲-based theories may be of the required order of magnitude but the values obtained are much too large.

However, first principles corrections appear, which make error factor 8 for (4a) and 4 for (4b) suspicious for 2 reasons.

(i) Just like in Bohr H theory, effective Coulomb attraction $-e^2/r_0$ at equilibrium is diminished by a repulsive term, exactly equal to $+\frac{1}{2}e^2/r_0$, as in Kratzer oscillator (3h). This first fundamental additive correction $(e^2/r_0)(-1+\frac{1}{2})=-\frac{1}{2}e^2/r_0$ of virial type, brings in correction factor $\frac{1}{2}$ at $r=r_0$. Although repulsive terms for vibrations rely on nucleon kinetic energies, invisible in BOA rearrangement (3b)-(3d), their effect must be taken into account to describe the equilibrium of the system at $r_0$.

(ii) A $2^d$ correction factor is *multiplicative instead of additive*. Both $\mathbf{H}_S$ and $\mathbf{H}_A$ contain terms for the 4 masses, securing these Hamiltonians use total dihydrogen mass $T=2m_H$. However, when the bond shows harmonic behavior, a single reduced mass[7] R for the dihydrogen oscillator appears, equal to $R=m_H^2/(m_H+m_H)=\frac{1}{2}m_H$, giving ratios of respectively $R/T=\frac{1}{4}$ and $R/m_H=\frac{1}{2}$. Since mass acts like a field scale factor, it is valid also for the Coulomb field in (3e) at all r. Therefore, first principles virial and reduced mass, both invisible in BOA (3b), generate correction factors of the required magnitude in an effortless way. Numerical correction factors $F_n$ equal to

$$F_n = \frac{1}{2}^n \qquad (4c)$$

with integer n $1 \leq n \leq 3$ can be used for both (4a) and (4b). BOA (3b) at $r=r_0$ with $F_2$ gives

$$F_2 S_{BOA} = S_{BOA}/4 = 78500 \text{ cm}^{-1} \text{ and } D_e = 39250 \text{ cm}^{-1} \qquad (4d)$$

very close indeed to observed splitting $S_{HH}$ (2d) and $H_2$ dissociation energy $D_e$ [16,18]. A similar result with $F_3$ applies for (4a) and more complex $S_\pm(r_0)$ [5] but leads to the same value (2d). Since QM cannot deal conclusively with $\underline{H}$, result (4d) is critical for QM, especially if simplicity were a valid criterion to judge on the merit of a theory for insoluble 4-particle systems like $H_2$. *Despite appearances, nearly exact result (4d) for Coulomb splitting at $r_0$ agrees with observation $S_{HH}$ (2d) for the dihydrogen bond. It validates the 2 simple bond theories (3a)-(3b), based on generic anti-symmetry, brought about by intra-atomic charge inversion. By extension, result (4d) provides with a huge molecular spectral $\underline{H}$-signature and proves that $\underline{H}$ occurs in nature, i.e. in the stable natural hydrogen molecule, usually but unjustly denoted by $H_2$.*

**V. Supporting evidence**

Since the ground state of a diatomic covalent bond is an anti-symmetric atom-antiatom pair

$$HH = [H\underline{H}; \underline{H}H] \qquad (5a)$$

as proved above, supporting evidence must be available.

(i) Solution (5a) complies with Pauli anti-symmetry for bound ground states.

---

[7] Bohr's $2\mu = 2mM/(m+M)$, with m and M electron and proton mass, is too small for dihydrogen vibrations.

(ii) Bonding in *covalent* $H_2$ (5a) obeys Coulomb's $e^2/r_0$ (4a)-(4c), i.e. *ionic bond energy*

$$D_{ion} = e^2/r_0 \qquad (5b)$$

If so, classical 19th century *ionic bonding model* (5b) resembles (5a), since it leads to

$$HH = [H\underline{H};\underline{H}H] \approx [H^+H^-;H^-H^+] \qquad (5c)$$

Here, an old *ionic charge-transfer mechanism* is replaced with a *modern intra-atomic charge-inversion mechanism*. Classical 19th century ionic views (5b) support (5a) by common sense [5,15].

(iii) Only *ionic bond energy* $D_{ion}$ (5b), not *covalent bond energy* $D_e$, unifies spectroscopic constants of all available *ionic and covalent* bonds between univalent atoms [5,6,19] (see also [15]). For dihydrogen, the analytical connection between $D_{ion}$ and $D_e$ is made explicit in [16].

(iv) Kratzer's (3h), a substitute for antiBOA (3d), retrieves the observed $H_2$ force constant $k_e = e^2/r_0^3 = 5{,}7 \cdot 10^5$ dyne/cm exactly as well as its 1st Dunham coefficient $a_0 = \tfrac{1}{2}k_e r_0^2 = 78000$ cm$^{-1}$ [15].

(v) As a result, Kratzer's (3h) also immediately retrieves, analytically, a fundamental frequency of dihydrogen equal to $\omega = 4390$ cm$^{-1}$ [15], where 4402 cm$^{-1}$ is observed [11,18]. These rather exact results with ionic Kratzer potential (3h), itself a substitute for antiBOA (3c) and therefore valid only for H$\underline{H}$ [15], fully support (5a).

(vi) The $H_2$ PEC, shown in Fig. 1 (full line), is more accurate than that of HL theory, since it is extracted directly from the observed vibrational levels [20]. This *experimental* curve is compared with the theoretical Kratzer PEC (3h) using solely $r_0 = 0{,}74$ Å [18] as input and theoretical BOA result $D_e = 4{,}75$ eV (4d) as well depth (dashed line). In line with (iv) and (v), the two curves nearly coincide not only around the minimum but even in about 90 % of the total well depth. Since Kratzer's (3h) is of antiBOA-type and refers to asymmetric H$\underline{H}$ instead of symmetric HH, Fig. 1 fully supports (5a). It certainly illustrates the effect of the conceptual simplicity of (3a)-(3b).

(vii) For *simple bond theories* (3a)-(3b) to make sense, complementary $\underline{H}$-signatures must show in the H line spectrum, which is exactly what we found [4]. Its Mexican hat curve [4], the basis of (1b), is typical for chiral systems with both H- and $\underline{H}$-states being bonding (see Section I).

(viii) A Hund-type Mexican hat curve is also found in the band spectrum of dihydrogen [16]. This confirms the presence of both H- and $\underline{H}$-states in the natural and stable dihydrogen bond as well. While the $H_2$ PEC in elaborate QM analysis [9] fails on these important aspects, we found that this curve is a *quartic of closed form*, without higher order terms needed [16].

(ix) For the 14 vibrational levels observed for dihydrogen [11], refined calculations on the basis of Kratzer's (3h) lead to errors of only 0,015 cm$^{-1}$ [16], even smaller than those in [9] (see Section III.2). Although our very precise results in [16] call for more accurate measurements of the $H_2$ band spectrum than hitherto [11], they first of all validate conclusion (5a).

This long list of cumulating supporting evidence is almost *incontournable* by its formal, conceptual and computational simplicity. Unlike [9], it is easily verified almost without calculations or wave functions but still produces an acceptable, reasonably accurate PEC for $H_2$. This makes it more difficult than ever to refute or to ignore this huge H-signature in the band spectrum of dihydrogen as large as 9,5 eV or 77000 cm$^{-1}$. If validated, the result has implications for H-theory as well as for H-experiments like [1,2].

**VI. Conclusion**

Claims [1,2], constrained by theory and conditioned by a complex experimental set-up[8], fail exactly where it really matters: *hard evidence for* H. In contrast, signatures for natural H [4-6] are clearly visible in simple spectra, available for a century, are understood with classical physics but are persistently ignored hitherto by those adhering to [1,2]. The largest ever H-signature of about 9,5 eV, reported here, is the observed symmetry-dependent splitting in the dihydrogen band spectrum [5]. A degree of freedom for charges in neutral species, instead of fixing charges by an a priori convention, makes the mysterious anti-world an intimate and integral part of the real world and turns the so-called matter-antimatter asymmetry of the Universe into a debatable issue [5].
A common sense, classical bond theory, allowing H, places question marks on QM bond theory, including the meaning of wave functions[9], on the BOA and on the theory behind [1,2]. With the LHC likely to be operational in 2008, new H-claims like [1,2] should be examined more critically.

---

[8] e.g. beams of e$^+$ and p$^-$, confinement, Penning-Ioffe traps with electrical and magnetic fields [3], particle *acceleration* followed by particle *deceleration*, cooling… and indirect H-detection with annihilation products.
[9] This H-result is critical for the concept of wave functions [5] (see [21] for references on this long-standing debate).

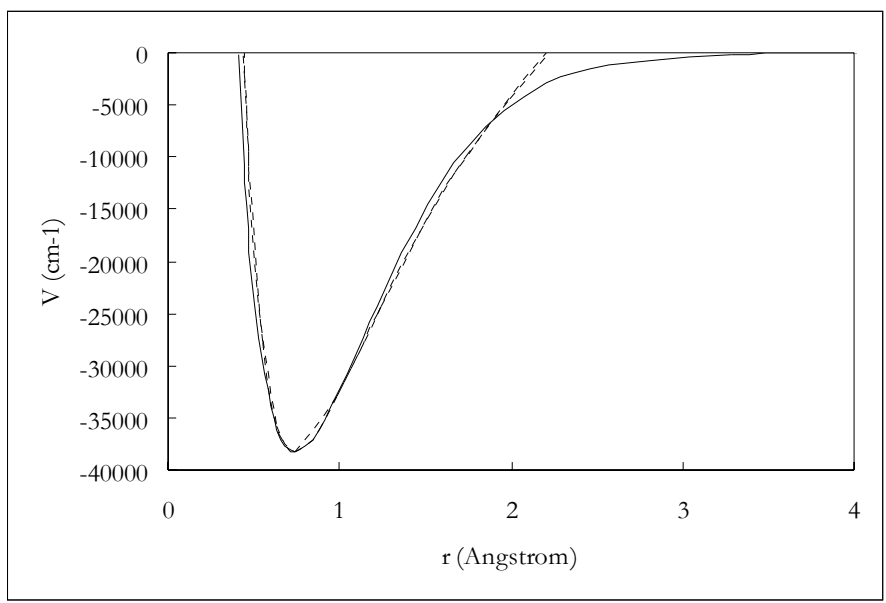

Fig.1 Observed [20] (full line) and theoretical Kratzer PEC (3h) (dashes) for $H_2$